\documentclass[pra,aps,groupaddress,prd]{revtex4}
\usepackage{amsfonts}
\usepackage{epsfig,amsmath,graphicx,amssymb,overpic,bm}
\usepackage{color,tikz,xcolor,extarrows}
\usetikzlibrary{arrows}

\usepackage{epsfig,amsmath,graphicx,amssymb,overpic}
\usepackage{bm}

\def\be{\begin{equation}}
\def\ee{\end{equation}}
\def\bee{\begin{eqnarray}}
\def\ene{\end{eqnarray}}
\def\bes{\begin{subequations}}
\def\ees{\end{subequations}}

\newcommand{\PT}{\mathcal{PT}}

\begin{document}

\baselineskip=13pt\title{Rogue wave formation and interactions in the defocusing nonlinear Schr\"odinger equation  with external potentials}
\author{Li Wang}
\author{Zhenya Yan}
\email{zyyan@mmrc.iss.ac.cn}
\affiliation{Key Lab of Mathematics Mechanization, Academy of Mathematics and Systems Science, Chinese Academy of Sciences, Beijing 100190, China\\
School of Mathematical Sciences, University of Chinese Academy of Sciences, Beijing 100049, China}

\vspace{0.1in}
\baselineskip=12pt

\begin{abstract} \vspace{0.05in}
{\bf Abstract.}\, The defocusing nonlinear Schr\"odinger (NLS) equation has no the modulational instability, and was not found to possess the rogue wave (RW) phenomenon up to now. In this paper, we firstly investigate some novel nonlinear wave structures in the defocusing NLS equation with real-valued time-dependent and time-independent potentials such that the stable new RWs and W-shaped solitons are found, respectively. Moreover, the interactions of two or three RWs are explored such that the RWs with higher amplitudes are generated in the defocusing NLS equation with real-valued time-dependent potentials. Finally, we study the defocusing NLS equation with complex $\PT$-symmetric potentials such that some RWs and W-shaped solitons are also found. These novel results will be useful to design the related physical experiments to generate the RW phenomena and W-shaped solitons in the case of defocusing nonlinear interactions, and to apply them in the related fields of nonlinear or even linear sciences.

\vspace{0.1in}
\end{abstract}

\maketitle

\baselineskip=12pt

\vspace{-0.3in}
\section{Introduction}

The rogue wave (RW) phenomenon should earliest be found in the deep ocean, and has caused many catastrophic damages due to its
high amplitudes and violent energy~\cite{Hopkin}. After the optical RWs was observed in the experiment, and play an important role in the  supercontinuum generation, the RWs have been paid more and more attention to. Up to now, the RWs have been theoretically or/and experimentally verified to appear in many distinct nonlinear physical systems such as Bose-Einstein condensates, fiber optical, plasma physics, and even financial markets~\cite{op}. The first analytic RW solution of the focusing NLS equation was found by Peregrine in 1983~\cite{ps}. In 2010, the Peregrine RW was verified to well agree with the numerical and experimental results nearby the origin in the focusing NLS equation~\cite{ps-t}. Though there exist many analytic, numerical, and experimental results about the RWs in the distinct nonlinear systems (see, e.g., Ref.~\cite{rw1} and references therein), the physical mechanisms generating RWs in nonlinear wave equations still needs to be investigated.

For the nonlinear wave equations, the modulational instability (MI) may be regarded as a necessary condition for generating the extreme waves (e.g., RW phenomenon). For example, the {\it focusing} NLS equation ($\sigma=-1$)
\begin{equation}
i u_{t}= - \frac{1}{2}u_{xx}+\sigma|u |^{2}u
\label{nlse}
\end{equation}
admits the MI, and was verified to possess the RWs by the analytical, numerical, and experimental ways. However, the {\it defocusing} NLS equation ($\sigma=1$)
has no MI~\cite{dnls}. Up to now, the defocusing NLS equation (\ref{nlse}) with $\sigma=1$ was not found to admit the RW phenomenon. A natural problem is what additional condition can make the defocusing NLS equation (\ref{nlse}) with $\sigma=1$ generate the RWs. In other word, what generalized form of the defocusing equation (\ref{nlse}) with $\sigma=1$ can admit the RWs ?

In this paper, we would like to consider the generalized forms of Eq.~(\ref{nlse}) with $\sigma=1$, i.e., the defocusing NLS equation with external potentials. We find that the defocusing equation with a time-dependent potential can support the stable RWs, and defocusing equation with an time-independent potential can support the stable W-shaped solitons. We verify that these nonlinear wave phenomena appearing in the focusing NLS equation can also be found in the generalized defocusing NLS equation with external potentials.
Recently, the $\PT$ symmetry plays a more and more important role in the linear spectral problems and nonlinear wave equations. In this paper, we will further consider the defocusing equation with $\PT$-symmetric potentials such that the RWs and W-shaped solitons can also be found.

The rest of this paper is arranged as follows. In Sec. 2, we consider the defocusing NLS equation with time-dependent and time-independent potentials such that the stable RWs and W-shaped solitons are found, respectively. Moreover, we study the interactions of two or three RWs in the time-dependent and time-independent potentials. In Sec. 3, we explore the generalized defocusing NLS equation with complex $\PT$-symmetric potentials such that the stable RWs and W-shaped solitons are also found. Moreover, we also analyze the effects of the gain-and-loss parameters in the $\PT$-symmetric linear spectral problems. Finally, some conclusions and discussions are given in Sec. 4.

\begin{figure}[!t]
\begin{center}
\vspace{0.05in} {\scalebox{0.45}[0.45]{\includegraphics{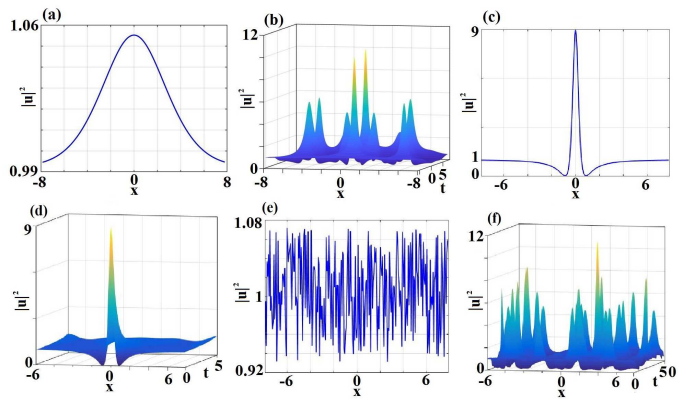}}}
\end{center}
\par
\vspace{-0.2in}
\caption{{\protect\small (color online). (a) The profile of RW (\ref{sol}) at $t=-6$ of the focusing NLS Eq.~(\ref{nlse}); (b) The wave evolution of initial condition (a) in the focusing NLS equation (\ref{nlse}); (c) The profile of RW (\ref{sol}) at $t=0$ of the focusing NLS Eq.~(\ref{nlse}); (d) The wave evolution of initial condition (c) in the focusing NLS equation (\ref{nlse}); (e) The profile of the plane wave at $t=0$ of the focusing NLS Eq.~(\ref{nlse}); (f) The wave evolution of initial condition (e) with a noise in the focusing NLS equation (\ref{nlse}).}}
\label{unstable_rw_standard}
\end{figure}

\section{Nonlinear Schr\"odinger equations and nonlinear waves}

\begin{figure*}[!t]
\begin{center}
\vspace{0.05in} {\scalebox{0.56}[0.56]{\includegraphics{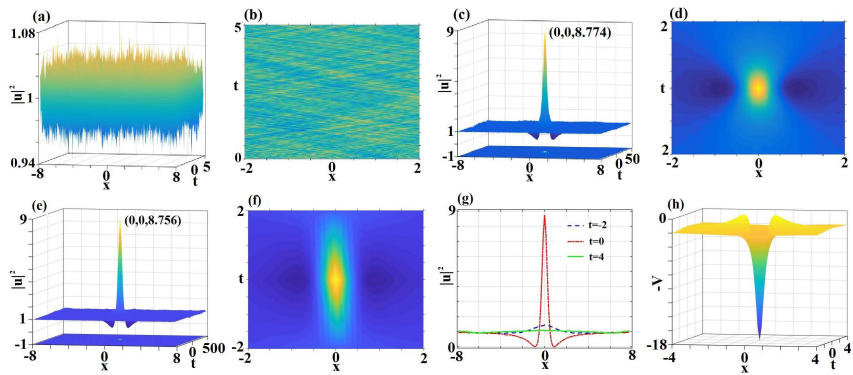}}}
\end{center}
\par
\vspace{-0.2in}
\caption{{\protect\small (color online). (a, b) The wave propagations of the plane wave at $t=0$ with a  random noise in the defocusing NLS Eq.~(\ref{nlse}) with $\sigma=1$; (c,d) The wave propagations of the plane wave at $t=0$ with a random nose in the defocusing NLS equation
with the time-dependent potential (\ref{nls}); (e,f) The evolution of RW (\ref{sol}) with $t=-500$ and a $2\%$ random noise in the defocusing NLS Eq.~(\ref{nls}) with the time-dependent potential (\ref{nls}); (g)  The intensity features in (e,f) for $t = -2$ (dashed blue line), $t = 0$ (dashed-dot red line), and $t = 4$ (solid green line); (h) The profile of time-dependent potential (\ref{nls}).} }
\label{vt}
\end{figure*}

{\it 2.1 \, The basic focusing and defocusing NLS equations.}---As a fundamental physical model, the {\it focusing} NLS equation (\ref{nlse}) appears in many fields of nonlinear science such as the nonlinear optics, Bose-Einstein condensate, deep ocean, plasmas physics, and even finance (see, e.g., Refs.~\cite{op, mal,prw9}), and was found to possess the basic RW solution~\cite{ps} (also called the Peregrine soliton or rogon~\cite{op,rw1})
\begin{equation}
u_s(x,t) = \left[ 1-\frac{4(1+2it)}{4(x^2+t^2)+1}\right] \mathrm{e}^{it},  \label{sol}
\end{equation}
from which one can have $|u_s(x,t)|\to 1$ as $|x|, |t|\to \infty$, ${\rm max}(|u_s(x,t)|)=3$ at the point $(x,t)=(0, 0)$, ${\rm min}(|u_s(x,t)|)=0$ at the points $(x,t)=(\pm \sqrt{3}/2, 0)$, and $u_s(x,t)\sim e^{it}$ (a plane wave) as $|x|, |t|\to \infty$. Moreover $P(t)=\int_{-\infty}^{\infty}|u_s(x,t)-e^{it}|^2dx=4\pi/\sqrt{4t^2+1}$ and $0<P(t)\leq 4\pi$. The Peregrine RW (\ref{sol})~\cite{ps} was originally derived from the parameter limit ($c_1\to 0$) of the temporal-periodic Kuznetsov-Ma breather~\cite{km1,km2} of Eq.~(\ref{nlse})
 \bee\label{km}
 u_{km}(x,t)\!=\!\frac{\cos(wt\!-\!2ic_1)\!-\!\cosh(c_1)\cosh(kx)}{\cos(wt)\!-\!\cosh(c_1)\cosh(kx)}e^{it},\hspace{-0.25in}
\ene
where $ w=\sinh(2c_1),\, k=2\sinh(c_1)$, and $c_1$ is a real constant. $|u_{km}(x,t)|^2\to 1$ and $u_{km}(x,t)\to e^{it}$ as $|x|\to \infty$.
In fact, it can also be deduced from the parameter limit ($c_2\to 0$) of spatial periodic Akhmediev breather~\cite{ab} of Eq.~(\ref{nlse})
  \bee \label{ab}
 u_{ab}(x,t)\!=\!\frac{\cosh(wt\!-\!2ic_2)\!-\!\cos(c_2)\cos(kx)}{\cosh(wt)\!-\!\cos(c_2)\cos(kx)}e^{it}, \hspace{-0.25in}
\ene
where $w=\sin(2c_2),\, k=2\sin(c_2)$, and $c_2$ is a real constant. $|u_{ab}(x,t)|^2\to 1$ and $u_{ab}(x,t)\to e^{it}$ as $|t|\to \infty$. Moreover, if we consider the complex $c_1$ and $c_1=ic_2$, then the temporal-periodic Kuznetsov-Ma breather (\ref{km}) becomes the spatial periodic Akhmediev breather (\ref{ab})~\cite{prw}.

We use $u_s(x, -6)$ and $u_s(x, 0)$ as the initial conditions (see Figs.~\ref{unstable_rw_standard}(a,c)) to study their wave evolutions in the focusing NLS equation (\ref{nlse}) such that we find that they are both unstable and generate many wave humps where there are some ones with larger amplitudes (see Figs.~\ref{unstable_rw_standard}(b,d)). In fact, Eq.~(\ref{nlse}) admits the modulational instability (MI), that is, one can use a simple plane-wave solution (e.g, $u=e^{it}$) with a $5\%$ random  noise as an initial condition (see Fig.~\ref{unstable_rw_standard}e) to possibly generate the RW phenomena in Eq.~(\ref{nlse}) (see Fig.~\ref{unstable_rw_standard}f).

However, the {\it defocusing} NLS equation  does not possess the MI (see Figs.~\ref{vt}(a,b), where we use the simple plane wave $u=e^{-it}$  with a $2\%$ random noise as an initial condition to study its wave evolution illustrating the modulational stability). Up to now, the defocusing NLS equation was not verified to support the RW phenomenon. A natural problem is that what condition can make the defocusing NLS equation support the RW phenomena.


\vspace{0.1in}

{\it 2.2. \, The defocusing NLS equation with external potentials and RWs.}---Nowadays, we consider the defocusing NLS equation with the real-valued time-dependent external potential in the dimensionless form
\bee
i u_{t}=-\frac{1}{2}u_{xx}-V(x,t)u+|u|^{2}u,\quad  V(x, t)=\frac{4(t^2-x^2)+1}{(x^2+t^2+0.25)^2}+2,
\label{nls}
\ene
where
$V$ approaches to $2$ as $|x|,|t|\to \infty$ and ${\rm max}(V(x,t))=18$ at the point $(x,t)=(0,0)$ (see Fig.~\ref{vt}h), $u(x,t)$ represents the wave envelope field with $t$ and $x$ being the propagation variable and transverse coordinates, respectively. Notice that the external potential (\ref{nls}) depends on both space and time.

More interesting, we find that the defocusing NLS equation with the time-dependent potential (\ref{nls}) also admits the analytical RW solution (\ref{sol}). This differs from the usual defocusing NLS equation (\ref{nlse}) without a potential, which was not found to admit the RWs up to now. That is to say, the external potential (\ref{nls}) can make the defocusing NLS equation (\ref{nls}) support the RWs. Particularly, we know that as $|x|, |t|\to\infty$, we have $u_s(x,t)\sim e^{it}$ and $V(x,t)\sim 2$.

\begin{figure}[t]
\begin{center}
\vspace{0.05in} {\scalebox{0.5}[0.5]{\includegraphics{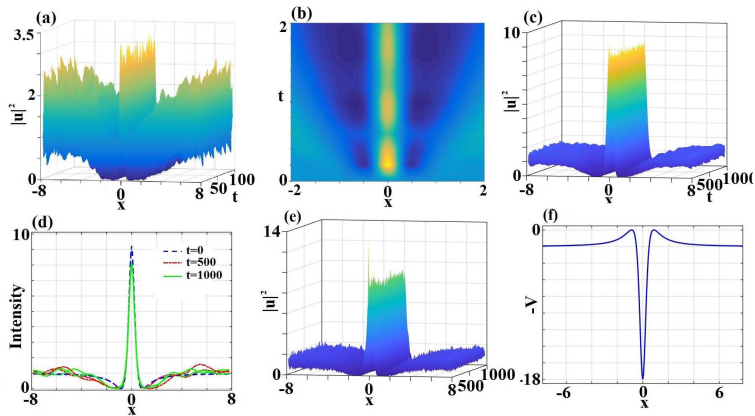}}}
\end{center}
\par
\vspace{-0.2in}
\caption{{\protect\small (color online).  (a, b)  The evolution of the plane wave $e^{-it}$ with $t=0$ and a $2\%$ random noise in the defocusing Eq. (\ref{nls}) and time-independent potential (\ref{gain0}); (c) The evolution of RW (\ref{sol}) with $t=0$ and $2\%$ random noise  in the defocusing Eq. (\ref{nls}) and time-independent potential (\ref{gain0}); (d) The intensity feature for $t = 0$ (dashed blue line), $t = 500$ (dashed-dot red line), and $t = 1000$ (solid green line); (e)  The evolution of RW (\ref{sol}) with $t=0$ and $40\%$ random noise  in the formation of Eq. (\ref{nls}) and time-independent potential (\ref{gain0}); (f)  The profile of time-independent potential (\ref{gain0}). } }
\label{v_pinned}
\end{figure}

\vspace{0.1in}
{\it Case 1.\, The time-dependent potential}: We now study the wave evolution of the simple plane wave $u=e^{-it}$ with $t=0$ and a $2\%$ random noise (i.e., $u(x,0)=e^{-0i}(1+2\%\times{\rm noise}))$ (see Fig.~\ref{vt}c) in the defocusing equation with the time-dependent potential (\ref{nls}) such that we can find the RW phenomenon (see Fig.~\ref{vt}d).
Next, we turn our attention to the stability of the RW solution (\ref{sol}) in the defocusing NLS equation (\ref{nls}) with the time-dependent potential (\ref{nls}). Figs.~\ref{vt}(e,f) illustrate the RW propagation in the defocusing NLS equation by using the RW (\ref{sol}) with $t=-500$ and a $2\%$ random noise (i.e., $u_s(x,-500)(1+2\%\times{\rm noise})$) as the initial condition such that we find that the defocusing NLS Eq.~(\ref{nls}) with the time-dependent potential (\ref{nls}) can almost allow for the stable existence of the RW formation. Moreover, we find that ${\rm max}(|u(x,t)|^2)\approx 8.756$ at the point $(0, 0)$. Fig.~\ref{vt}(g) exhibits the RW profiles at the distinct times $t=-2, 0, 4$.

These above-mentioned  results imply that the defocusing NLS equation  with the time-dependent potential (\ref{nls}) can admit the RW phenomena (see Figs.~\ref{vt}(c,e)), and the time-dependent potential (\ref{nls}) plays a key role in the RW formation even if the defocusing NLS equation (\ref{nls}) without an external  potential $(i.e., V(x,t)=0)$ does not possess the RW (see Fig.~\ref{vt}b). In other words, the RWs can also be stably formed under the interplay between the {\it time-dependent potential} (\ref{nls}) and {\it defocusing} Kerr nonlinearity. The result about the RWs was found in the generalized defocusing NLS equation before.

\vspace{0.1in} {\it Case 2.\, The time-independent potential}: We now consider the effect of time-independent potential  given  by Eq.~(\ref{nls}) with $t=0$, that is,
 \bee
V_0(x)=V(x,0)=2\left(\frac{x^2-0.75}{x^2+0.25}\right)^{\!2},  \label{gain0}
\ene
which approaches to $2$ as $x\to \pm\infty$ (see Fig.~\ref{v_pinned}f), in the defocusing NLS equation (\ref{nls}).

Similarly, we consider the wave evolution of the simple plane wave $u(x,t)=e^{-it}$ with $t=0$ and a $2\%$ random noise in the defocusing equation with the time-independent potential (\ref{gain0}) such that we also can find the W-shaped soliton (see Figs.~\ref{v_pinned}(a,b)). We still use the RW solution (\ref{sol}) with $t=0$ and a $2\%$ random noise (i.e., $u_s(x, 0)(1+2\%\times {\rm noise}$)) as the initial condition to study the wave propagation in the defocusing NLS equation (\ref{nls}) with the time-independent potential (\ref{gain0}) such that the stable W-shaped soliton (not RW) is found (see Figs.~\ref{v_pinned}(c,d)). Moreover, It is easy to see that even if the larger $40\%$ random noise can not damage the stable evolution of the W-shaped soliton (see Fig.~\ref{v_pinned}e).

It follows from Figs.~\ref{vt} and \ref{v_pinned} that for the same defocusing NLS equation (\ref{nls}) and similar initial conditions of RWs, the time-dependent potential (\ref{nls}) can support the stable RW (see Fig.~\ref{vt}e), however the time-independent potential (\ref{gain0}) can support the stable W-shaped solitons (see Fig.~\ref{v_pinned}c).


\begin{figure}[!t]
\begin{center}
\vspace{0.05in} {\scalebox{0.6}[0.6]{\includegraphics{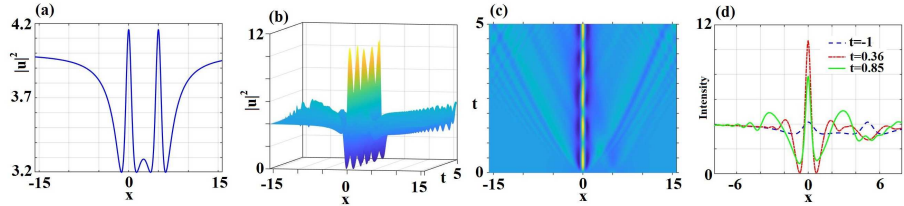}}}
\end{center}
\par
\vspace{-0.25in}
\caption{\small (color online). The interaction of solutions in the defocusing Eq.~(\ref{nls}) with the time-independent potential (\ref{gain0}).
 (a) The initial condition is $u_s(x,1)+u_s(x-5,1)$; (b, c) The interaction of two RWs; (d) The profiles at $t =-1$ (dashed blue line), $t = 0.36$ (dashed-dot red line), and $t = 0.85$ (solid green line).}
\label{coll_2W}
\end{figure}

\vspace{0.1in}
{\it 2.3. \, The interactions of RWs in time-dependent and time-independent potentials.}---It is known that the dissipative term was proposed generally to investigate the RW formation~\cite{Lecaplain}. Moreover, the higher-order perturbation terms could also excite the RW generations in the focusing NLS equation~\cite{Ankiewicz, Bandelow, yanchaos15, yanaml19,yanchaos2020}.

{\it Case 1.\, The time-independent potential}: We now consider the interactions of RWs in the defocusing Eq. (\ref{nls}) and time-independent potential (\ref{gain0}). Figs.~\ref{coll_2W}(b,c,d) display the interaction between two RWs $u_s(x,1)$ and $u_s(x-5,1)$ (we use $u_s(x,1)+u_s(x-5,1)$ as the initial condition, see Fig.~\ref{coll_2W}(a)) in Eq.~(\ref{nls}) with the time-independent potential (\ref{gain0}). As a result, we find that the interaction can generate the W-shaped soliton with higher amplitudes and semi-periodic oscillations. Moreover, the intensity of W-shaped soliton almost approaches to $4$ as $|x|\to \infty$.

\begin{figure}[!t]
\begin{center}
\vspace{0.05in} {\scalebox{0.6}[0.6]{\includegraphics{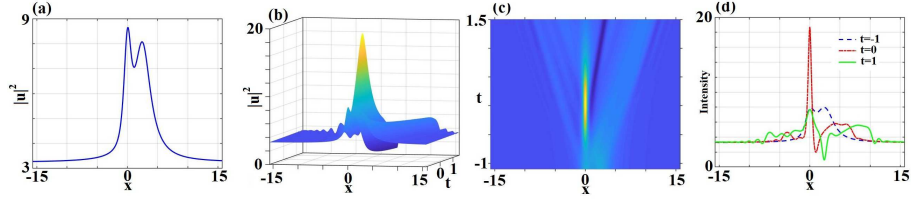}}}
\end{center}
\par
\vspace{-0.25in}
\caption{\small (color online).  The interaction of solutions in the defocusing equation with the time-dependent potential (\ref{nls}).
 (a) The initial condition is $u_s(x,0.9)+u_r(x-2.5,0.9)$ with $u_r(x,t)=u_s(x,t)+1-e^{it}$; (b, c) The interaction of two RWs; (d) The profiles at $t =-1$ (dashed blue line), $t = 0$ (dashed-dot red line), and $t =1$ (solid green line).}
\label{coll_2RW}
\end{figure}

\begin{figure}[!t]
\begin{center}
\vspace{0.05in} {\scalebox{0.6}[0.6]{\includegraphics{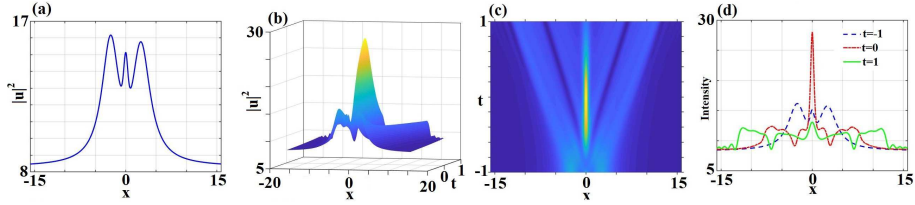}}}
\end{center}
\par
\vspace{-0.25in}
\caption{\small (color online).  The interaction of solutions in the defocusing equation with the time-dependent potential (\ref{nls}).
 (a) The initial condition is $u_s(x,0.9)+u_r(x-2.5,1.9)+u_r(x+2.5,1.8)$ with $u_{r}(x,t)=u_s(x,t)+1-e^{it}$; (b, c) The interaction of three RWs; (d) The profiles at $t =-1$ (dashed blue line), $t = 0$ (dashed-dot red line), and $t =1$ (solid green line).}
\label{coll_3RW}
\end{figure}

{\it Case 2.\, The time-dependent potential}: We now consider the interactions of RWs in the defocusing equation and time-dependent potential given by Eq.~(\ref{nls}).  Figs.~\ref{coll_2RW}(b,c,d) display the interaction between two RWs $u_s(x,0.9)$ and $u_r(x-2.5,0.9)$
with $u_r(x,t)=u_s(x,t)+1-e^{it}$ (we use $u_s(x,0.9)+u_r(x-2.5,0.9)$ as the initial condition, see Fig.~\ref{coll_2RW}(a)) in Eq.~(\ref{nls}) with the time-dependent potential (\ref{nls}). As a result, we find that the interaction can generate the RW with higher amplitudes. Furthermore, we also display the interaction of three RWs. Figs.~\ref{coll_3RW}(b,c,d) display the interaction between two RWs $u_s(x,0.9)$ and $u_r(x-2.5,1.9)$, and $u_r(x+2.5,1.8)$ (we use $u_s(x,0.9)+u_r(x-2.5,1.9)+u_r(x+2.5,1.8)$ as the initial condition, see Fig.~\ref{coll_3RW}(a)) in Eq.~(\ref{nls}) with the time-dependent potential (\ref{nls}). As a result, we find that the interaction can generate the RW with higher amplitude than one in  Fig.~\ref{coll_2RW} for the interaction of two RWs.

\section{The defocusing NLS equation with $\mathcal{PT}$-symmetric potentials}

In the section, we consider the defocusing equation with both the external potential (\ref{nls}) and gain-and-loss distribution $W(x)$ in the form
\begin{eqnarray}
i u_{t}= - \frac{1}{2}u_{xx}-[V(x,t)+iW(x)]u+|u|^{2}u,   \label{mnls}
\end{eqnarray}
where $W(x)$ is an odd function of space, i.e., $W(x)=-W(-x)$. It is easy to know that the complex potential $U(x, t)=-[V(x,t)+iW(x)]$ is $\mathcal{PT}$-symmetric~\cite{Bender1}, i.e., $U(x,t)=U^*(-x, t)$.

Here we choose the following three kinds of basic gain-and-loss distributions $W(x)$ as~\cite{pt1,pt2,pt3,pt4,pt5}
\bee\label{tan}
W_1(x) = W_{0}\tanh x, \quad W_2(x)=W_{0}\,\mathrm{sech} x\tanh x, \quad
W_3(x) = W_{0} x e^{-x^{2}},
\ene
where $W_0$ is a real constant.
We find that $W_1\to \pm W_0$ as $x\to\pm \infty$, while $W_{2,3}\to \pm 0$ as $x\to\pm \infty$ and $W_3$ decreases more quickly than $W_2$. That is, the gain-and-loss distribution $W_j(x)$ has the weaker and weaker effect on the defocusing equation (\ref{mnls}) as $j$ becomes large.


\begin{figure*}[!t]
\begin{center}
\vspace{0.05in} {\scalebox{0.5}[0.5]{\includegraphics{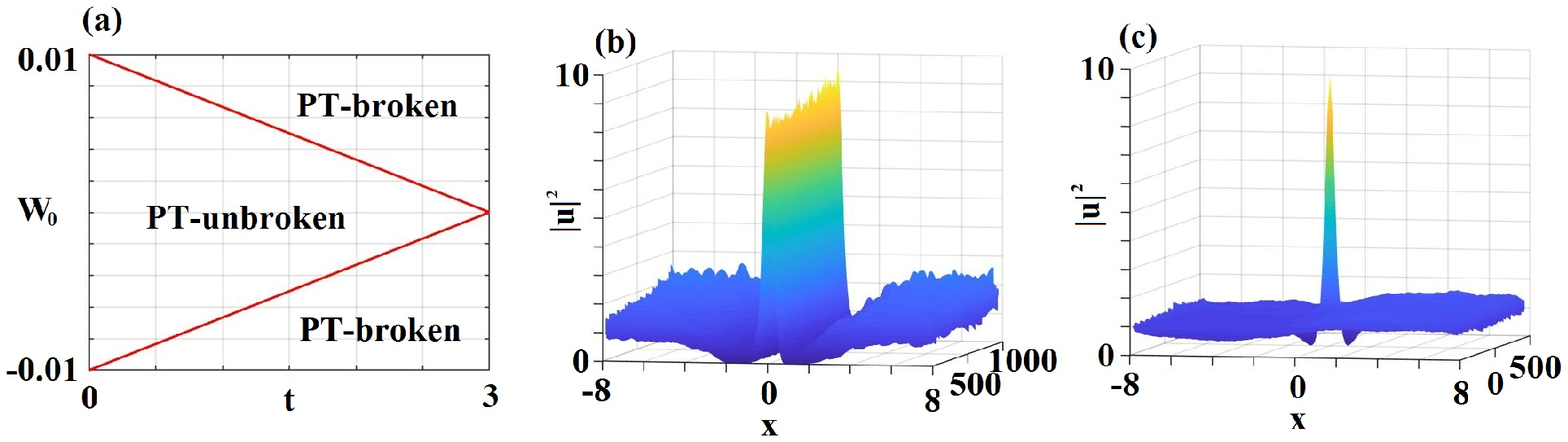}}}
\end{center}
\par
\vspace{-0.3in}
\caption{{\protect\small (color online).  (a) The region of $\PT$-broken/unbroken phases for $V(x,t)$ and $W_1(x)$ given by Eqs.~(\ref{nls}) and (\ref{tan}) in the ($t$, $W_{0})$ space; (b) The stable W-shaped soliton evolution of initial condition given by Eq.~(\ref{sol}) with $t=0$ and a 2$\%$ random noise in the framework of the defocusing NLS equation (\ref{mnls}), the time-independent potential (\ref{gain0})  and gain-and-loss distribution (\ref{tan}) with $W_0=0.04$; (c)  The stable RW evolution of initial condition given by Eq.~(\ref{sol}) with $t=-500$ and a 2$\%$ random noise in the framework of the defocusing NLS equation (\ref{mnls}), the time-dependent potential (\ref{nls}), and gain-and-loss distribution (\ref{tan}) with $W_0=0.02$. }}
\label{pt_tanh}
\end{figure*}

\begin{figure}[!t]
\begin{center}
\vspace{0.05in} {\scalebox{0.6}[0.6]{\includegraphics{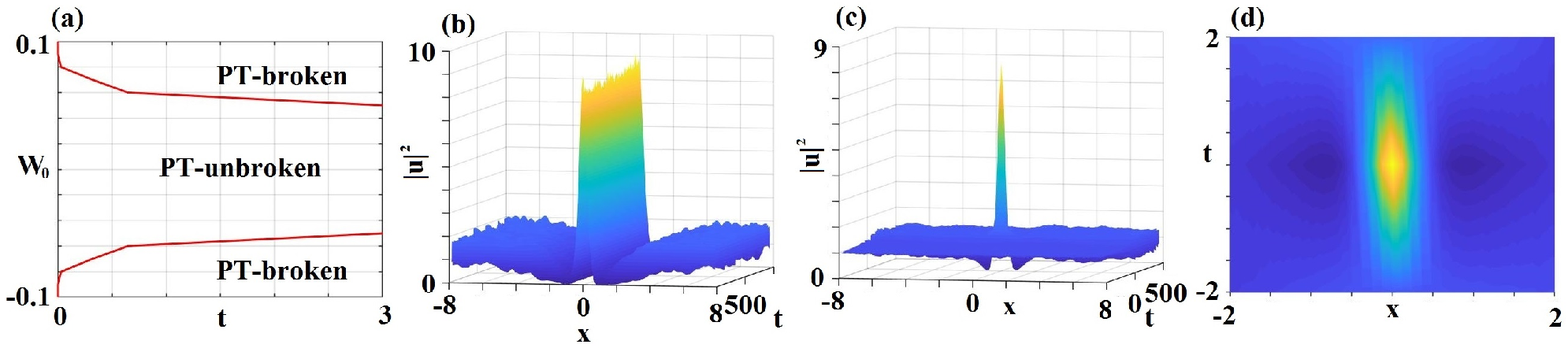}}}
\end{center}
\par
\vspace{-0.2in}
\caption{{\protect\small (color online).   (a) The region of $\PT$-broken/unbroken phases for $V(x,t)$ and $W_2(x)$ given by Eqs.~(\ref{nls}) and (\ref{tan}) in the ($t$, $W_{0})$ space; (b) The stable W-shaped soliton evolution of initial condition given by Eq.~(\ref{sol}) with $t=0$ and a 2$\%$ random noise in the framework of the defocusing NLS equation (\ref{mnls}), the time-independent potential (\ref{gain0})  and gain-and-loss distribution (\ref{tan}) with $W_0=0.9$; (c, d)  The stable RW evolution of initial condition given by Eq.~(\ref{sol}) with $t=-500$ and a 2$\%$ random noise in the framework of the defocusing NLS equation (\ref{mnls}), the time-dependent potential (\ref{nls}), and gain-and-loss distribution (\ref{tan}) with $W_0=0.06$. } }
\label{pt_scarff}
\end{figure}

\begin{figure}[!t]
\begin{center}
\vspace{0.05in} {\scalebox{0.6}[0.6]{\includegraphics{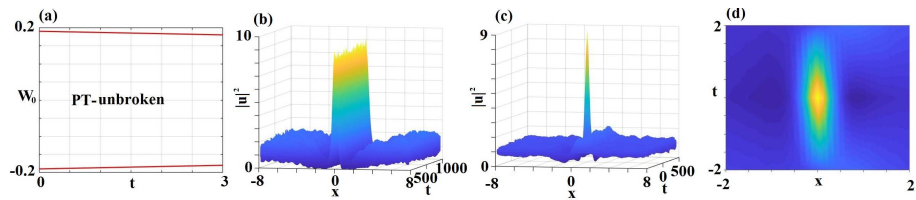}}}
\end{center}
\par
\vspace{-0.2in}
\caption{{\protect\small  (color online).(a) The region of $\PT$-broken/unbroken phases for $V(x,t)$ and $W_3(x)$ given by Eqs.~(\ref{nls}) and (\ref{tan}) in the ($t$, $W_{0})$ space; (b) The stable W-shaped soliton evolution of initial condition given by Eq.~(\ref{sol}) with $t=0$ and a 2$\%$ random noise in the framework of the defocusing NLS equation (\ref{mnls}), the time-independent potential (\ref{gain0})  and gain-and-loss distribution (\ref{tan}) with $W_0=2.98$; (c, d) The stable RW evolution of initial condition given by Eq.~(\ref{sol}) with $t=-500$ and a 2$\%$ random noise in the framework of the defocusing NLS equation (\ref{mnls}), the time-dependent potential (\ref{nls}), and gain-and-loss distribution (\ref{tan}) with $W_0=2.1$. } }
\label{pt_gauss}
\end{figure}

We firstly consider the linear $\PT$-symmetric spectral problem~\cite{Bender1}:
$\mathcal{L}\psi= \lambda \psi, \quad \mathcal{L}=-\frac{1}{2} \partial_{x}^{2} - V(x,t) - iW(x)$,
where $V(x,t)$ is given by Eq.~(\ref{nls}), $W(x)$ is chosen as $W_j(x)$ given by Eq.~(\ref{tan}), $\lambda$  stands for eigenvalue and $\psi$ is the underlying eigenvector, $\mathcal{L}$ is $\PT$ symmetric, and  $t$ can be treated as a parameter, not a variable.

Since the $\mathcal{PT}$-symmetric property may provide possibility to admit real eigenenergies for the non-Hermitian $\PT$ Hamiltonian $\mathcal{L}$. 
The real and complex eigenvalue distributions related to  $\mathcal{L}$ associated with Eqs.~(\ref{nls}) and (\ref{tan}) are illustrated in Fig.~\ref{pt_tanh}(a) by virtue of numerical Fourier spectral method. It is easy to see that there
always exists a threshold (i.e., two red lines): the real spectra (i.e., $\PT$-unbroken phases) exist inside the two lines, however the complex spectra (i.e., $\PT$-broken phases) exist beyond the two red lines. The intriguing finding is that for the point $(t, W_0)=(0, 0.04)$ in Fig.~\ref{pt_tanh}(a) where the linear spectral problem has the {\it complex eigenvalues} and the time-dependent potential $V(x,t)$ given by Eq.~(\ref{nls}) reduces to the time-independent potential $V_0(x)$ given by Eq.~(\ref{gain0}), we use Eq.~(\ref{sol}) with $t=0$ and a 2$\%$ random noise as an initial condition to study its evolution such that the stable W-shaped soliton can exist in Eq.~(\ref{mnls}) in the presence of  the time-independent potential (\ref{gain0}) and gain-and-loss distribution (\ref{tan}) with $W_0=0.04$ (see Fig.~\ref{pt_tanh}b). That is to say, the nonlinear term may broaden the linear $\mathcal{PT}$-symmetric threshold allowing  real  eigenvalues. For the time-dependent potential (\ref{nls}),
we use Eq.~(\ref{sol}) with $t=-500$ and a 2$\%$ random noise as an initial condition to study its evolution in Eq.~(\ref{mnls}) in the presence of  the time-dependent potential (\ref{nls}) and gain-and-loss distribution (\ref{tan}) with $W_0=0.02$ such that the stable RW can be found (see Fig.~\ref{pt_tanh}c).

Similarly, we consider the linear spectral problem with $V(x,t)$ and $W_2(x)$ or $W_3(x)$ such that the thresholds about $\PT$-unbroken/broken phases are given in Figs.~\ref{pt_scarff}a and \ref{pt_gauss}a, respectively. It follows from Figs.~\ref{pt_tanh}a, \ref{pt_scarff}a and \ref{pt_gauss}a that for the given potential $V(x,t)$, if the gain-and-loss distribution $W(x)$ has the weaker effect on the
linear spectral problem, then the region for the $\PT$-unbroken phases becomes the larger. Fig.~\ref{pt_scarff}b implies that when we use Eq.~(\ref{sol}) with $t=0$ and a 2$\%$ random noise as an initial condition to study its evolution such that the stable W-shaped soliton can exist in Eq.~(\ref{mnls}) under the sense of the time-independent potential (\ref{gain0}) and gain-and-loss distribution $W_2$ with $W_0=0.9$ (see Fig.~\ref{pt_scarff}b). For the time-dependent potential (\ref{nls}),
we use Eq.~(\ref{sol}) with $t=-500$ and a 2$\%$ random noise as an initial condition to study its evolution in Eq.~(\ref{mnls}) under the sense of the time-dependent potential (\ref{nls}) and gain-and-loss distribution $W_2$ with $W_0=0.06$ such that the stable RW can be found (see Figs.~\ref{pt_scarff}(c,d)). For the given gain-and-loss distribution $W_3$, we have also the similar results. Fig.~\ref{pt_gauss}b implies that when we use Eq.~(\ref{sol}) with $t=0$ and a 2$\%$ random noise as an initial condition to study its evolution such that the stable W-shaped soliton can exist in Eq.~(\ref{mnls}) under the sense of the time-independent potential (\ref{gain0}) and gain-and-loss distribution $W_3$ with $W_0=2.98$ (see Fig.~\ref{pt_scarff}b). For the time-dependent potential (\ref{nls}), we use Eq.~(\ref{sol}) with $t=-500$ and a 2$\%$ random noise as an initial condition to study its evolution in Eq.~(\ref{mnls}) under the sense of the time-dependent potential (\ref{nls}) and gain-and-loss distribution $W_3$ with $W_0=2.1$ such that the stable RW can be found (see Figs.~\ref{pt_gauss}(c,d)).

Figures~\ref{pt_scarff}b and \ref{pt_gauss}b show the evolution of stable W-soliton with $2\%$ random noise, even though the corresponding pinned values of $t = 0$ , and $W_{0} = 0.9$ and $2.98$, respectively, in Figs.~\ref{pt_scarff}a and \ref{pt_gauss}a are associated with complex eigenenergies, leading to eigenenergies oscillating.  These results reported in Figs.~\ref{pt_scarff} and \ref{pt_gauss} are similarly to ones
in Fig.~\ref{pt_tanh}. However, the thresholds presented in Figs.~\ref{pt_tanh}a, \ref{pt_scarff}a and \ref{pt_gauss}a allowing for the unbroken $\mathcal{PT}$-symmetry eigenstates become bigger for three different dissipative terms given by Eq.~(\ref{tan}),
and the maximal dissipative coefficient subject to evolution of stable RWs becomes bigger (see Fig.~\ref{pt_tanh} (c) with $W_{0} = 0.02$, Fig.~\ref{pt_scarff}c with $W_{0} = 0.06$ and Fig.~\ref{pt_gauss}c with $W_{0} = 2.1$). We would like further to investigate the characteristics produced by  real external potential and Gaussian dissipative term $W_3$ in Fig.~\ref{pt_gauss}. Figs.~\ref{pt_gauss}b and  \ref{pt_gauss}c correspond to the case, the impact of gain-and-loss system satisfying the $\mathcal{PT}$ symmetry is required to simulate Hamiltonian dynamics in the defocusing NLS equation (\ref{mnls}).

In fact the threshold condition of the parameter $\mathrm{W_{0}}$ is $W_0\in [0, 2.1]$ to make sure the defocusing NLS equation (\ref{mnls}) with the time-dependent potential possess the stable RWs, whereas the threshold condition of the parameter $\mathrm{W_{0}}$ in $W_3$ is $W_0\in [0, 2.98]$ to make sure the defocusing NLS equation (\ref{mnls}) with the time-independent potential admit a stable W-soliton. Obviously,
we find that the larger the value of $\mathrm{W_{0}}$ is, the higher the background amplitudes become during the study of the numerical simulations.


In conclusion, we have verified that the defocusing NLS equation (\ref{nls}) with the time-dependent potential (\ref{nls}) could support the analytical RWs, and numerically found that the RWs could stably exist in the defocusing model with the time-dependent potential. Moreover, we find that the defocusing NLS equation (\ref{nls}) with the time-independent potential (\ref{gain0}) can support the stable W-shaped solitons. We also study the interactions of  two or three RWs in the defocusing NLS equation (\ref{nls}) with time-dependent or time-independent potentials such that the W-shaped solitons and RWs with higher amplitudes are found. Finally, we explore the defocusing NLS equation (\ref{mnls}) with some $\PT$-symmetric potentials. As a consequence, the stable W-shaped solitons and RWs can be generated again.



These above-mentioned results imply that the external potentials play an significant role in the study of the defocusing NLS equation and even linear Schr\"odinger equation. The idea used in this paper can also be extended to other nonlinear equations without MI such that these models with some potentials may generate the RW phenomena. We will further study these questions in future. \\



\noindent{\bf Acknowledgments}\, The work was supported by the NSFC under Grant Nos. 11731014 and 11925108, and CAS Interdisciplinary Innovation Team.



\end{document}